\documentclass[12pt,preprint]{aastex}

\slugcomment{paper draft to ApJ (ver. 07/04/2005)}

\shorttitle{Mapping Observations of IRAS 19312+1950}
\shortauthors{Nakashima \& Deguchi}

\begin{document}


\title{BIMA Array Observations of the Highly Unusual SiO Maser Source with a Bipolar Nebulosity, IRAS 19312+1950}


\author{Jun-ichi Nakashima\altaffilmark{1}}
\affil{Department of Astronomy, University of Illinois at Urbana-Champaign,
1002 W. Green St., Urbana, IL 61801}
\email{junichi@asiaa.sinica.edu.tw}

\and

\author{Shuji Deguchi}
\affil{Nobeyama Radio Observatory, National Astronomical Observatory, Minamimaki, Minamisaku, Nagano 384-1305, JAPAN}
\email{deguchi@nro.nao.ac.jp}

\altaffiltext{1}{Current Address: Academia Sinica Institute of Astronomy and Astrophysics, P.O. Box 23-141, Taipei 106, Taiwan}


\begin{abstract}
We report the results of mapping observations of the bipolar nebula with SiO maser emission, IRAS 19312+1950, in the CO ($J=1$--0 and $J=2$--1), $^{13}$CO ($J=1$--0 and $J=2$--1), C$^{18}$O ($J=1$--0), CS ($J=2$--1), SO ($J_{\rm K}=3_{2}$--2$_{1}$) and HCO$^{+}$ ($J=3$--2) lines with the Berkeley-Illinois-Maryland Association array. Evolutional status of this source has been evoking a controversy since its discovery, though SiO maser sources are usually identified as late-type stars with active mass loss. In line profiles, two kinematical components are found as reported in previous single-dish observations: a broad pedestal component and a narrow component. Spatio-kinetic properties of a broad component region traced by $^{12}$CO lines are roughly explained by a simple spherical outflow model with a typical expanding velocity of an AGB star, though some properties of the broad component region still conflict with properties of a typical AGB spherical outflow. A narrow component region apparently exhibits a bipolar flow. The angular size of the narrow component region is spatially larger than that of a broad component region. Intensity distribution of the CS emission avoids the central region of the source, and that of an SO broad component emission exhibits a small feature peaked exactly at the mapping center. According to the present results, if a broad component really originates in a spherical outflow, an oxygen-rich evolved stellar object seems to be a natural interpretation for the central star of IRAS 19312+1950.
\end{abstract}




\begin{abstract}
We report the results of mapping observations of the bipolar nebula with SiO maser emission, IRAS 19312+1950, in the CO ($J=1$--0 and $J=2$--1), $^{13}$CO ($J=1$--0 and $J=2$--1), C$^{18}$O ($J=1$--0), CS ($J=2$--1), SO ($J_{\rm K}=3_{2}$--2$_{1}$) and HCO$^{+}$ ($J=3$--2) lines with the Berkeley-Illinois-Maryland Association array. Evolutional status of this source has been evoking a controversy since its discovery, though SiO maser sources are usually identified as late-type stars with active mass loss. In line profiles, two kinematical components are found: a broad pedestal component and a narrow component. Spatio-kinetic properties of a broad component region are roughly explained by a simple spherical outflow model with a typical expanding velocity of an AGB star, though some properties still conflict with those of a typical AGB spherical outflow. A narrow component region apparently exhibits a bipolar flow. The angular size of the narrow component region is spatially larger than that of a broad component region. Intensity distribution of the CS emission avoids the central region of the source, and that of an SO broad component emission exhibits a small feature peaked exactly at the mapping center. According to the present results, if a broad component really originates in a spherical outflow, an oxygen-rich evolved stellar object seems to be a natural interpretation for the central star of IRAS 19312+1950.
\end{abstract}




\section{Introduction}
Evolutional status of the SiO maser source, IRAS 19312+1950, has been evoking a controversy since its discovery in our SiO maser survey toward a sample of relatively cold IRAS sources \citep{nak00,nak03}. Infrared colors and morphology of the source are fairly reminiscent of those of asymptotic giant branch (AGB) stars (or post-AGB stars); in fact, SiO maser sources are usually identified as oxygen-rich (O-rich) late-type stars with active mass loss \cite[see, e.g.,][]{hab96}. Some properties of the source, however, are consistent with those of young stellar objects (YSOs) in molecular clouds \citep{nak00,nak04a,nak04b,deg04}. IRAS 19312+1950 exhibits a twin-peaked SiO maser line profile \citep{nak00}; this exhibits strong time variation \citep{deg04}. The twin-peaked SiO maser profile is rarely seen in AGB stars, and is reminiscent of the SiO maser profile of the Orion KL star-forming region \citep{sny74,mor77,pla90,doe99}. H$_{2}$O (22.2351 GHz) and OH (1612, 1665, and 1667 MHz) maser lines have also been detected \citep[][Lewis 2000 in private communications]{nak00}. The H$_{2}$O maser line profile shows several intensity peaks with strong time variation \citep{deg04}. The OH maser profiles show complicated spiky profiles, which are unusual in AGB stars, but the line intensity ratios between three different OH maser transitions are consistent with those of type II OH maser sources in AGB stars \citep[i.e., the 1612 MHz satellite line is the strongest among the three lines, see, e.g.,][]{hab96}.

A rich set of molecular species, detected toward IRAS 19312+1950, has recently raised an exotic possibility that this source is an AGB/post-AGB star lying in a dark cloud \citep{deg04}, though the idea is not conclusive yet. In fact, 16 molecular species (including isotope species) have been detected so far toward the source \citep{deg04}; this richness of molecular species is reminiscent of that of OH 231.8+4.2 \citep[rotten egg nebula, see, e.g.,][]{mor87}. The molecular line profiles often show two kinematic components: strong lines with small line-widths of $\sim$ 2 km s$^{-1}$ (narrow component) and weak-symmetric lines with large line-widths of $\gtrsim$ 60 km s$^{-1}$ (broad pedestal component) \citep{nak04b,nak04a}. The broad pedestal component seems to be attributed to an AGB wind in terms of its large line-width, but the narrow component is rather difficult to explain simply by an AGB/post-AGB star origin \citep{nak04a}. Such a narrow line profile is relatively rare but known in several late-type stars \citep[for example, EP Aqr, X Her, EU And, and BM Gem;][]{kna98,ker99,jur99,win03,nak05}. A recent interferometric radio observation of IRAS 19312+1950 in the HCO$^{+}$ $J=1$--0 line \citep{nak04a} has revealed existence of a bipolar flow and a (conceivably) spherical outflow with a small angular size. If IRAS 19312+1950 is a true AGB/post-AGB star, as suggested in our previous papers, the rich molecular environment of the source offers a compelling opportunity to investigate the chemical environment of a unique O-rich AGB/post-AGB envelope. Otherwise, even if the source identified as a YSO, it must provide an alternative opportunity to investigate a YSO with SiO masers, because the number of YSOs, emitting SiO masers, are very limited. However, further astrophysical interpretation of the source is contingent upon secure identification of its evolutional status. 

In this paper, we present the results of a set of high resolution radio interferometric observations in 8 molecular rotational lines with the Berkeley-Illinois-Maryland Association (BIMA) array. Main objectives in the present research are (1) to reveal a precise spatio-kinetic properties of narrow and broad component regions, and (2) to investigate chemical properties of the source, with high-angular-resolution radio mapping observations. We make use of the present results to consider the evolutional status of the source. The outline of this paper is as follows: in section 2, details of observations and results are presented. In section 3, spatio-kinetic properties of narrow and broad component regions are discussed;  chemical properties of the source are also briefly discussed in this section. Finally, results are summarized in section 4.


\section{Observations and Results}
Interferometric observations of IRAS 19312+1950 were made with the BIMA array; the instrument has been described in detail by \citet{wel96}. We observed the CO ($J=1$--0 and $J=2$--1), $^{13}$CO ($J=1$--0 and $J=2$--1), C$^{18}$O ($J=1$--0), CS ($J=2$--1), SO ($J_{\rm K}=3_{2}$--2$_{1}$) and HCO$^{+}$ ($J=3$--2) lines in two observing periods: February -- June 2003 (for CO) and December 2003 -- January 2004 (for CS, SO and HCO$^{+}$). Rest frequencies of the lines were taken from \citet{lov92}. The observations were interleaved every 25 minutes with the nearby point source, 1925+211, to track the phase variations over time. The absolute flux calibration was determined from observations of Uranus and MWC349, and is accurate to within 20\%. Typical single sideband system temperatures ranged from 200 K to 300 K (at $\lambda \sim$3 mm) and from 700 K to 1500 K ($\lambda \sim$1 mm). The velocity coverage was about 380 km s$^{-1}$ ($\lambda \sim$3 mm) and 150 km s$^{-1}$ ($\lambda \sim$1 mm), using three different correlator windows with a bandwidth of 50 MHz each. The velocity resolution was about 1 km s$^{-1}$ ($\lambda \sim$3 mm) and about 0.5 km s$^{-1}$ ($\lambda \sim$1 mm). The phase center of the maps is 19$^{\rm h}$33$^{\rm m}$24.4$^{\rm s}$, 19$^{\circ}$56$'$54.8$''$ (J2000) corresponding to the IRAS position of the target. Data reduction was performed with the MIRIAD software package \citep{sau95}. Standard data reduction, calibration, imaging, and deconvolution procedures were followed. The rms noises, integration times on the source, beam sizes given by robust weighting of the visibility data, position angles of the beams, and array configurations are summarized in Table 1. Note that about 50\% of the visibility data of the CO $J=1$--0 line observations were flagged out because of severe mechanical problems on a correlation system; therefore, the integration time of the CO $J=1$--0 line observations is practically a half of the time indicated in Table 1.

\subsection{Spectra}

In Figures 1 we present spectra of molecular lines observed, averaged spatially over a circle with an angular diameter of 15$''$; this averaging circle is centered on the mapping center. The line profiles, shown in Figure 1, are rather similar with those taken in previous single-dish observations with careful background subtraction \citep{nak04b,deg04}; this suggests that almost all of the single-dish flux is recovered by the present interferometric observations. In comparison with the previous single-dish data taken by the Nobeyama 45-m telescope \citep{nak04b}, about 70\% of flux, detected by the Nobeyama 45-m telescope, is recovered in the present interferometric observations; the main reason of the difference of flux is due presumably to difference of calibration methods in each telescope. The line profiles exhibit two kinematical component, as reported by \citet{nak04b}: ``narrow component'' peaked strongly at $V_{\rm lsr}=$38--40 km s$^{-1}$ and ``broad pedestal component''. The narrow component is seen in all lines observed, but the broad component appears to be limited in the CO $J=1$--0, CO $J=2$--1, $^{13}$CO $J=2$--1, and SO $J_{\rm K}=3_{2}$--2$_{1}$ lines (the broad component is particularly prominent in the CO $J=1$--0 and CO $J=2$--1 lines). The $^{13}$CO $J=1$--0 and CS $J=2$--1 lines seem to have a weak shoulder at the blueshifted side of the narrow component. According to intensity maps, the shoulder feature of the $^{13}$CO $J=1$--0 line seems to be a part of the broad pedestal component, whereas that of CS $J=2$--1 seems to be a part of the narrow component (i.e., narrow and broad components show different spatio-kinetic properties; we discuss this in later sections). Our previous observation in the HCO$^{+}$ $J=1$--0 line shows a broad pedestal component with a parabolic profile \citep{nak04a}; however, the present observation in the HCO$^{+}$ $J=3$--2 very marginally shows a weak broad component. This is due presumably to high noise level at the frequency very close to an edge of band coverage. Peak velocities of narrow components of the CO $J=1$--0 and CO $J=2$--1 lines are lying at $\sim$ 40 km s$^{-1}$, while narrow components of other lines are peaked at $\sim$38 km s$^{-1}$. This disagreement of peak velocities is due to strong absorption features lying at $V_{\rm lsr} \sim$35 and $\sim$45 km s$^{-1}$. A single-dish spectrum in the CO $J=1$--0 with no background subtraction \citep[see, Fig. 1 in ][]{nak04b} shows a strong emission feature at $V_{\rm lsr} \sim$35 km s$^{-1}$; therefore, the absorption feature at $V_{\rm lsr} \sim$35 km s$^{-1}$ is most likely real absorption due to foreground material, detected by the single-dish observation as an emission line. Although no strong emission was detected at $V_{\rm lsr} \sim$45 km s$^{-1}$ in the single-dish observation, the 45 km s$^{-1}$ feature also seems to be attributed to foreground clouds. This is because, as we will see in channel--velocity maps later, red- and blueshifted sides of the absorption feature lying at $V_{\rm lsr} \sim$45 km s$^{-1}$ exhibit almost same spatio-kinematic properties. The $^{13}$CO $J=1$--0 and C$^{18}$O $J=1$--0 lines, as well as the $^{12}$CO lines, seem to exhibit weak absorption features. \citet{nak04b} suggested, on the basis of their single-dish data, that a full width of the broad pedestal component of the CO $J=1$--0 line reaches, at least, to 60 km s$^{-1}$; in fact, the present interferometric data show that the full-width at the zero intensity level reaches nearly to 80 km s$^{-1}$. The CO $J=2$--1, $^{13}$CO $J=2$--1 and HCO$^{+}$ $J=3$--2 lines were first observed toward IRAS 19312+1950 in the present observations. Intensity of the CO $J=2$--1 and $^{13}$CO $J=2$--1 lines is clearly larger than that of the CO $J=1$--0 and $^{13}$CO $J=1$--0 lines. Approximate intensity ratios between higher and lower transitions of CO lines are [CO ($J=2$--1)]:[CO ($J=1$--0)] $=$ 1.76:1 (narrow component), [CO ($J=2$--1)]:[CO ($J=1$--0)] $=$ 8.5:1 (broad component), [$^{13}$CO ($J=2$--1)]:[$^{13}$CO ($J=1$--0)] $=$ 6.7:1 (narrow component). The intensity ratios of narrow components were calculated with peak intensity. The ratios of broad components were calculated with integrated intensity, which was integrated over particular velocity ranges (i.e., 15--20 km s$^{-1}$). The ratio of a broad component of the $^{13}$CO lines is not given here, because that is too weak to measure fluxes in reliable accuracy. In the C$^{18}$O $J=1$--0 line profile, a weak emission-like feature is seen at $V_{\rm lsr} \sim$ 110 km s$^{-1}$. We verified, with integrated intensity maps, that the possible high velocity component at $V_{\rm lsr} \sim$ 110 km s$^{-1}$ is not real. 

Continuum emission was not detected at a 3 $\sigma$ rms level both in 1 mm and 3 mm bands; to confirm the continuum emission, frequency channels were integrated over 150 MHz using the opposite-side bands to line observations. Upper limits of the continuum radiation flux at 3 mm and 1 mm bands (center frequencies of integration ranges are $\sim$107 GHz and $\sim$228 GHz) are $2.2 \times 10^{-3}$ Jy and 4.9 $\times 10^{-2}$ Jy, respectively.

\subsection{Intensity Maps}

Figures 2 and 3 show velocity channel maps of the $^{13}$CO $J=1$--0 and $^{13}$CO $J=2$--1 lines, respectively. The structure seen in Figures 2 and 3 is clearly resolved by our synthesized beam. Broad and narrow components mentioned in section 2.1 exhibit clearly different spatial structure in the velocity channel maps; this difference is prominent especially in Figure 3. In Figure 3, the velocity channels of 35--37 km s$^{-1}$ roughly correspond to a narrow component, and those of 24--32 and 40--46 km s$^{-1}$ correspond to a broad component. Broad component features appear to be roughly spherical morphology, and the sizes of the broad component features are smaller than those of narrow component features. The narrow component features extends in the north-west to south-east direction. In Figure 2, a narrow component is predominantly seen in velocity channels of 29--30 km s$^{-1}$ and 35--38 km s$^{-1}$. No strong features are seen in channels of 32--34 km s$^{-1}$ due to absorption in the corresponding velocity range. Broad component features are somewhat difficult to see in Figure 2 because of its weak intensity; however, a small feature peaked at the mapping center is seen in the 40 km s$^{-1}$ channel. In both of Figures 2 and 3, narrow component features appear to exhibit systematic variation as a function of radial velocity: narrow component features symmetrically vary with respect to the velocity $\sim$36.5 km s$^{-1}$. In maps of a blueshifted side (i.e., 35 and 36 km s$^{-1}$ channels), narrow component features extend mainly to the north-west direction, and exhibit almost no extension to the south-east direction; in contrast, features at a redshifted side (37 and 38 km s$^{-1}$ channels) extend mainly to the south-east direction, and almost no extension is seen to the north-west direction. Similar variation of structure is seen also in velocity channel maps of the C$^{18}$O emission; the maps are presented in Figure 4.

In Figures 5 and 6 we present velocity selected channel maps of the CO $J=1$--0 and CO $J=2$--1 lines, respectively. As well as the $^{13}$CO and C$^{18}$O lines, spherical and elongated features are seen: velocity channels of 37--39 km s$^{-1}$, corresponding to a narrow component, exhibit largely extended features elongated in the north-west to south-east direction, and those of $V_{\rm lsr}<26$ km s$^{-1}$ and $V_{\rm lsr}>42$ km s$^{-1}$, corresponding to a broad pedestal component, exhibit small sizes and apparent sphericity. A narrow component region of the CO $J=1$--0 emission is about 2 times larger than that of the CO $J=2$--1 emission; in contrast, a broad component region of the CO $J=1$--0 emission appears to be about 2 times smaller than that of the CO $J=2$--1 emission, although signal-to-noise ratios in the CO $J=1$--0 observations are somewhat small because of significant elimination of bad data as mentioned in the previous section. The systematic change of structure with respect to $V_{\rm lsr} \sim $36.5 km s$^{-1}$, seen in the $^{13}$CO and C$^{18}$O lines, is not seen in the CO $J=1$--0 and CO $J=2$--1 lines due presumably to interruption of absorption features.

To see overall intensity distribution of broad and narrow components, in Figures 7 and 8 we present velocity integrated intensify maps of each kinematic component. Figure 7 shows integrated intensity maps of the narrow components. To clarify a systematic change of structure with respect to $V_{\rm lsr} \sim 36.5$ km s$^{-1}$, we integrated intensity of narrow components over two different velocity ranges: 35--36 and 37--38 km s$^{-1}$. In Figure 7 intensity maps of these integration ranges are indicated by the thick (35--36 km s$^{-1}$) and thin (37--38 km s$^{-1}$) contours. The thick and thin contours exhibit clear difference in their maps; this difference is prominent in the $^{13}$CO $J=1$--0, $^{13}$CO $J=2$--1 and C$^{18}$O $J=1$--0 lines. At a glance, features composed by thin and thick contours are reminiscent of a bipolar flow. In the case of $^{13}$CO, a feature extends in the north-west to south-east direction, and has an angular size of about 50$''$ $\times$ 30$''$ with a position angle of about 130$^{\circ}$. These angular sizes correspond to linear sizes of $1.5 \times 10^{18}$ and $9.0 \times 10^{17}$ cm at 2 kpc; similarly, $3.0 \times 10^{18}$ and $1.8 \times 10^{18}$ cm at 4 kpc \citep[see, e.g., ][on distance estimation to the source]{nak04b}. Position angles of major axes of the elongated features seem to be slightly varies from line to line. If this feature is of a bipolar out-flow, the north-west side should be closer to us, and the south-east side should be further from us. In the cases of $^{12}$CO $J=1$--0 and $J=2$--1 lines, emission in the range of $V_{\rm lsr}=35$--36 km s$^{-1}$ is almost totally obscured by absorption features; therefore, features are almost absent in thick contours. Although features in Figure 7 are mainly elongated in the north-west to south-east direction, another minor elongation is seen in the south-west side of the mapping center especially in maps of the $^{13}$CO $J=1$--0 and C$^{18}$O $J=1$--0 lines; near-infrared structure \citep[see, e.g.][]{nak04a} is also elongated in this direction. Note that intensity peak positions of the $^{13}$CO and C$^{18}$O narrow components are lying at 3$''$ north-west of the center, and do not correspond to the mapping center,.

In Figure 8 we present velocity integrated intensity maps of broad components. The right and left columns represent maps of blue- and redshifted wings of broad components, respectively. In addition to CO maps, in Figure 8 a similar map of the SO line is also presented. Features seen in Figure 8 are clearly resolved by our synthesized beam except for SO, and morphology of features are totally different from that of narrow components. At a glance, structure seen in Figure 8 seems to be roughly spherical. Intensity peak positions of red- and blueshifted wings mutually correspond within a beam size. The angular diameter of the $^{12}$CO (largest one) $J=2$--1 spherical feature is about 11$''$, corresponding to $3.3 \times 10^{17}$ at 2 kpc (or $6.6 \times 10^{17}$ cm at 4 kpc). In the case of SO, the structure is not resolved by our synthesized beam; this suggests that angular sizes of the SO broad component regions is less than about 3$''$.

In Figure 9, we present velocity integrated intensity maps of narrow components of CS, SO and HCO$^{+}$. A map of the HCO$^{+}$ $J=1$--0 line is also presented for comparison; the HCO$^{+}$ $J=1$--0 data were taken from \cite{nak04a}. The intensity maps are superimposed on a near-infrared $J$, $H$, $K$--bands composite image. In Figure 9, similarly with Figure 7, the thick and thin contours represent velocity integration ranges of 34--36 and 37--39 km s$^{-1}$. Note that scales of the maps are different from those of CO maps (i.e., the region of Figure 9 are about 6 times smaller than that of Figures 2--5). In contrast to the CO lines, no significant structure variation with respect to radial velocity is seen in the CS, SO and HCO$^{+}$ lines, and structure of these lines is clearly different from that of the CO line. Features seen in Figure 9 appear to avoid the mapping center, where near-infrared emission exhibits a bright peak. A notable intensity peak is lying at about 3$''$ north-west of the mapping center; this position is common in all four lines, and corresponds with intensity peak positions of the $^{13}$CO and C$^{18}$O narrow components. As mentioned in the section 2.1, the CS line profile shows a weak shoulder at the redshifted side of its peak; this shoulder component shows an intensity peak at the position 3$''$ north-west of the mapping center. The SO map exhibits another weak peak at about 2$''$ south-east of the mapping center; HCO$^{+}$ ($J=3$--2) exhibits a peak at about 7$''$ east of the mapping center. A notable point in Figure 9 is asymmetry of intensity distributions: all lines appears to be stronger at the north-west side than at the south-east side of the mapping center. This asymmetry seems to show significant difference of physical condition between the north-west and south-east sides.


\section{Discussion}
In this section, we have discussed evolutional status of IRAS 19312+1905 and related problems by making use of the present results.

\subsection{Properties of the Broad Component Region}

A prompt interpretation of a broad component region, apparently exhibiting spherical morphology in intensity maps, would be a spherically expanding outflow. For better understanding of spatio-kinetic properties of the broad component region, in Figure 10 we present a structure size vs. radial velocity diagram using the CO $J=1$--0, CO $J=2$--1 and $^{13}$CO $J=2$--1 data. Though the SO line also exhibits a weak broad component, here we do not use the SO data, because structure of the SO broad component region is not resolved by our synthesized beam. In Figure 10 structure sizes are represented by 5 $\sigma$ contours indicated in Figures 3, 5 and 6. We measured the sizes in the north--south and east--west directions, and then the values measured in the two orthogonal directions were averaged; a half of this average was adopted as angular sizes in Figure 10. The averages should correspond to radii if features are spherical. One might think that sizes given by least square fitting by a circle is an useful alternative. In fact the circle fitting gives roughly similar results, but occasionally produces unreliable values due to irregularity of features; therefore, we do not adopt here the circle fitting. The error bars in the vertical axes represent a half of differences between measures in the two orthogonal directions, representing degree of deviation from a true circle. Observational data points are superimposed on simple models of spherical outflows \citep[see, e.g.,][]{fon03} with different outflow velocities. Model curves are fitted to the data by a least square; free parameters of the models are only a radius and systemic velocity. At a glance, data points of CO $J=1$--0 and CO $J=2$--1 appear to be fitted well by a model with an expansion velocity of 30 km s$^{-1}$. Although, in the case of CO $J=2$--1, data points in higher velocity parts (i.e., $-$2 and 67 km s$^{-1}$ channels) somewhat deviate from the model curve of $V_{\rm exp}=30$ km s$^{-1}$, this is presumably because the features in higher velocity ranges are not resolved by our synthesized beam. 

If a spherical outflow is the case, the broad component region should be identified as an AGB wind, because spherical outflows are intrinsically harbored only in late-type stars, but not in YSOs. With the expansion velocity of $V_{\rm exp}=30$ km s$^{-1}$, a dynamical scale of a broad component region is estimated to be between 1600 and 3100 yr (on the assumption of distance 2--4 kpc), which is valid for a time scale of an AGB circumstellar envelope. However, several problems still remains in the spherical outflow interpretation. First, if the broad components really originates in a spherically expanding outflow, a line profile should be a parabolic or flat top shape \citep[a profile of a optically thin line in a spherically expanding shell is flat-topped, and that of a thick line is parabolic when the telescope beam cannot resolve the envelope,][]{mor77a}; however, the line profile of broad components of the $^{12}$CO $J=2$--1 line looks to be different from both of these profiles, although broad components of other lines seem to show parabolic shapes. In fact, the $^{12}$CO $J=2$--1 line profile looks a Gaussian profile rather than parabolic or flat top profiles. Second, the angular size of a broad component region is larger in the CO $J=2$--1 line than in the CO $J=1$--0 line; this is very unusual if the broad component originates in a usual spherical AGB envelope. To explain this, outermost part of the broad component region should be somehow heated to enhance the CO $J=2$--1 intensity. To explain these problems, shock between a central spherical flow and ambient material might play a key role. If the shock region have relatively large turbulent velocity, the region might produce a high-velocity wing in spectra, possibly exhibiting a Gaussian-like profile; this region might also exhibit relatively high temperature due to the shock, causing the larger size of the CO $J=2$--1 emission. If this is to the point, the highest velocity part of the $^{12}$CO $J=2$--1 line should exhibit in a shell-like feature in intensity maps. Unfortunately, we cannot confirm such a feature due conceivably to insufficient integration time. As a future work, higher sensitivity observations, mapping weak wings, would be important.

Third, properties of $^{13}$CO $J=2$--1 (see bottom panel in Figure 10) are significantly different from those of CO $J=1$--0 and CO $J=2$--1. In fact, the best fit model of a spherical outflow to the $^{13}$CO $J=2$--1 data gives $V_{\rm exp} \sim 10$ km s$^{-1}$ and $V_{\rm peak} \sim 36$ km s$^{-1}$; these values are largely different from those of CO $J=1$--0 and CO $J=2$--1. One possible explanation for this difference is that a $^{13}$CO broad component feature is affected by absorption. In fact, velocity ranges of strong absorption features seen in $^{12}$CO spectra are overlapped with those of the $^{13}$CO broad component. Although the $^{13}$CO line is relatively optically thin compared with the $^{12}$CO line, the $^{13}$CO line might be still affected by absorption if its optical depth is very large. Another possible explanation for the $^{13}$CO properties is that a spherical outflow has a shell structure expanding spherically, and the inside of the shell exhibits different kinematics with the shell (for example, a bipolar flow). Actually, detached shell structure has been found in post-AGB stars \cite[see, e.g.,][]{uet05}. However, in such a case, large number density of molecules, which enable us to detect the $^{13}$CO line, is not expected in the hollow of the shell, because post-AGB stars usually do not exhibit significant mass loss. This explanation using the hollow structure also conflicts with detection of SiO masers, because SiO masers are usually emitted from a near-stellar region. 

Another notable point of a broad component is an intensity ratio of broad component emission of the SO and SiO (thermal) lines. According to spectra in \citet{deg04}, broad components of the SO and SiO (thermal) lines have about the same strength; however, in an O-rich AGB envelope, the SiO thermal line is expected to be substantially stronger than the SO line. Though we did not observed the SiO thermal line in the present research, the SiO line is important as a future work to clarify this problem. At the present stage, in our opinion, the spherical outflow interpretation is not conclusive, but also it cannot be irrevocably ruled out.

\subsection{Properties of the Narrow Component Region}

A rich set of molecular species of IRAS 19312+1950 is one of the central scientific problems of this source, because the number of detected molecular species is unexpectedly large as an O-rich AGB/post-AGB star, and also because some species  detected toward the source (i.e., HCN, HNC, CN, CS, HC$_{3}$N, and methanol) are rarely found in O-rich AGB/post-AGB stars \citep[see,][]{deg04}. With results in the present observations, the central star most likely has O-rich chemistry, for reasons as follows. First, a broad component is found in the SO line, but not found in that of the CS line; this indicates that SO molecules exist in a central, spherical wind, but CS molecules does not. Second, the broad component of SO clearly exhibits an intensity peak at the mapping center (corresponding to the intensity peak of near-infrared emission); this suggests that the SO broad component originates in the central star. Third, the intensity distribution of the CS emission avoids the central region. If we rely on the central O-rich chemistry, C-bearing molecules, such as HCN and CN, are conceivably lying in outer regions, corresponding to a narrow component region. In fact, narrow components in molecular rotational spectra of IRAS 19312+1950 have raised difficulty in evolutional status identification of the source, and have been discussed repeatedly in our previous papers \citep{nak04b,nak04a,deg04}. 

As seen in Figure 7, a narrow component region apparently shows symmetric structure elongated in the north-west to south-east direction; this structure appears to exhibit a systematic motion. For better understanding of kinematical properties of the narrow component region, we made position--velocity maps in two orthogonal cuts; one corresponds to an apparent major axis and another is mutually perpendicular to it (cuts are indicated in Figure 7). Position angles of the cuts were determined by eye inspection. In Figure 11 we present the position--velocity maps. Systematic variation of structure as a function of radial velocity is clearly seen in cuts A; this is particularly prominent in the $^{13}$CO $J=1$--0 line, exhibiting a feature symmetrically placed with respect to $V_{\rm lsr} \sim 36.5$ km s$^{-1}$. In the maps of CO $J=1$--0 and $J=2$--1, no features are seen in red-shifted sides (i.e., $V_{\rm lsr} <36.5$ km s$^{-1}$); this is due to absorption in corresponding velocity ranges. Features of C$^{18}$O $J=1$--0 and $^{13}$CO $J=1$--0 seen in cuts B appear to also exhibit a systematic variation; other lines do not exhibits any strong systematic variation in cuts B. In our opinion, the systematic variation seen in Figure 11 is explained mainly by a bipolar flow rather than a Keplerian rotation disk, because the C$^{18}$O $J=1$--0 and $^{13}$CO $J=1$--0 lines exhibits velocity structure variation also in cuts B. Though weak rotational motion, coexisting with a bipolar flow, cannot be fully ruled out, the rotational motion would not be dominant. Note that \citet{nak04a} have also reported a bipolar flow, detected in the HCO$^{+}$ $J=1$--0 line; they suggested that the bipolar axis is lying in the north-east to south-west direction, which is perpendicular to the direction of a bipolar flow found in the present observations in the CO lines. We note, here, some differences of properties between bipolar flows detected in the HCO$^{+}$ and CO observations. First, a size of the CO bipolar flow is about 3 times larger than that of the HCO$^{+}$ bipolar flow; this means the HCO$^{+}$ bipolar flow is lying relatively closer to the central star compared with the CO bipolar flow. Second, an outflow velocity of the CO bipolar flow ($\sim$ 2 km s$^{-1}$) is 10 times smaller than that of the HCO$^{+}$ bipolar flow ($\sim$ 20 km s$^{-1}$); this potentially means that inclination angles of these bipolar flows are different if the flows have a same outflow velocity. These differences suggest that kinematical properties of a circumstellar envelope of IRAS 19312+1950 alter with distance from the central star. Otherwise, two distinct bipolar flows, which have different scale and different outflow velocities, might be individually lying in the envelope (i.e., multiple bipolar flows).

A relatively large size of a narrow component region compared with a size of a broad component region causes a rather difficult problem if we assume that all material has been expelled from a single central star. Apparently, a narrow component region is several times larger than a broad component region; this potentially means that a dynamical time scale of the bipolar flow is larger than that of the central spherical flow, because an outflow velocity of the bipolar flow is roughly 20 times smaller than that of the central spherical flow (of course, this depends on the inclination angle of the bipolar flow). AGB researchers widely believe that AGB circumstellar envelopes evolve from a spherical flow to a bipolar flow (though some post-AGB stars remain in spherical); the reverse case of this (i.e. from bipolar to spherical) has been not expected so far both from theoretical and observational view points. One possible solution of this problem is to assume a binary system consisting of an AGB and a post-AGB stars, such as OH 231.8+4.2 (rotten egg nebula). OH 231.8+4.2 is believed to be a binary system including two evolved stars in different evolutional stages: an AGB star (mira type variable with SiO masers) and a post-AGB star (exhibiting a bipolar molecular flow) \citep{san04}. In fact, IRAS 19312+1950 is fairy resemble with OH 231.8+4.2 in many aspects: a rich set of molecular species, detection of masers and a largely extended bipolar flow. Nevertheless, large luminosity expected in a binary system causes a problem again: if the binary system is the case, absolute luminosity of the source, at least, would be twice of the single-stellar luminosity. A larger luminosity places the object further in distance, forcing envelope physical parameters more extreme as an AGB star \citep[see, e.g., ][]{deg04, nak04b}. Besides, if the binary system includes an AGB star, the AGB star would most likely exhibit time variation in infrared luminosity \citep[see, e.g.,][]{nak00b}; however, recent near-infrared monitoring found that IRAS 19312+1950 does not show large variation in near-infrared, though small luminosity variation cannot be denied (Fujii et al. 2004, in private communications). To pursue this problem, more reliable distance is required; SiO masers, detected toward the source, may be helpful to determine the distance by measuring the annual parallax if we use a high-sensitive VLBI array.


\section{Summary}

In this paper, we have reported results of high-angular-resolution radio observations of the highly unusual SiO maser source, IRAS 19312+1950, in the CO ($J=1$--0 and $J=2$--1), $^{13}$CO ($J=1$--0 and $J=2$--1), C$^{18}$O ($J=1$--0), CS ($J=2$--1), SO ($J_{\rm K}=3_{2}$--2$_{1}$) and HCO$^{+}$ ($J=3$--2) lines with the BIMA array. Main results are as follows:
\begin{enumerate}
\item In line profiles, two kinematical components are seen as reported by previous single-dish observations: a broad pedestal component and narrow component. 

\item A broad component region, traced by the $^{12}$CO line, apparently exhibits spherical properties with expanding velocity $V_{\rm lsr} \sim 30$ km s$^{-1}$; this suggests that a broad component originates in an AGB wind. However, several problems still remain; for example, a Gaussian-like line profile is seen in the $^{12}$CO $J=2$--1 line; the $^{12}$CO $J=2$--1 emission is more spatially extended than the $^{12}$CO $J=1$--0 emission.

\item A narrow component region traced by CO lines apparently exhibits a bipolar flow, which is spatially larger than a broad component region. The dynamical time scale of the bipolar flow might larger than that of a possible spherical flow if we observe the source nearly in the pole-on direction, causing a conflict with an interpretation by a single AGB star.

\item Intensity distribution of the CS line avoids the central region of the source, and that of a broad component of SO exhibits a small feature, not resolved by our synthesized beam, exactly at the mapping center; this means that the central star has O-rich chemistry, and also that C-bearing molecules are conceivably lying at the outer part of the nebulosity.
\end{enumerate}


\acknowledgments
This research has been supported by the Laboratory for Astronomical Imaging at the University of Illinois and NSF grant AST 0228953. This research has made use of the SIMBAD and ADS databases.

\clearpage


\begin{figure}
\epsscale{.40}
\plotone{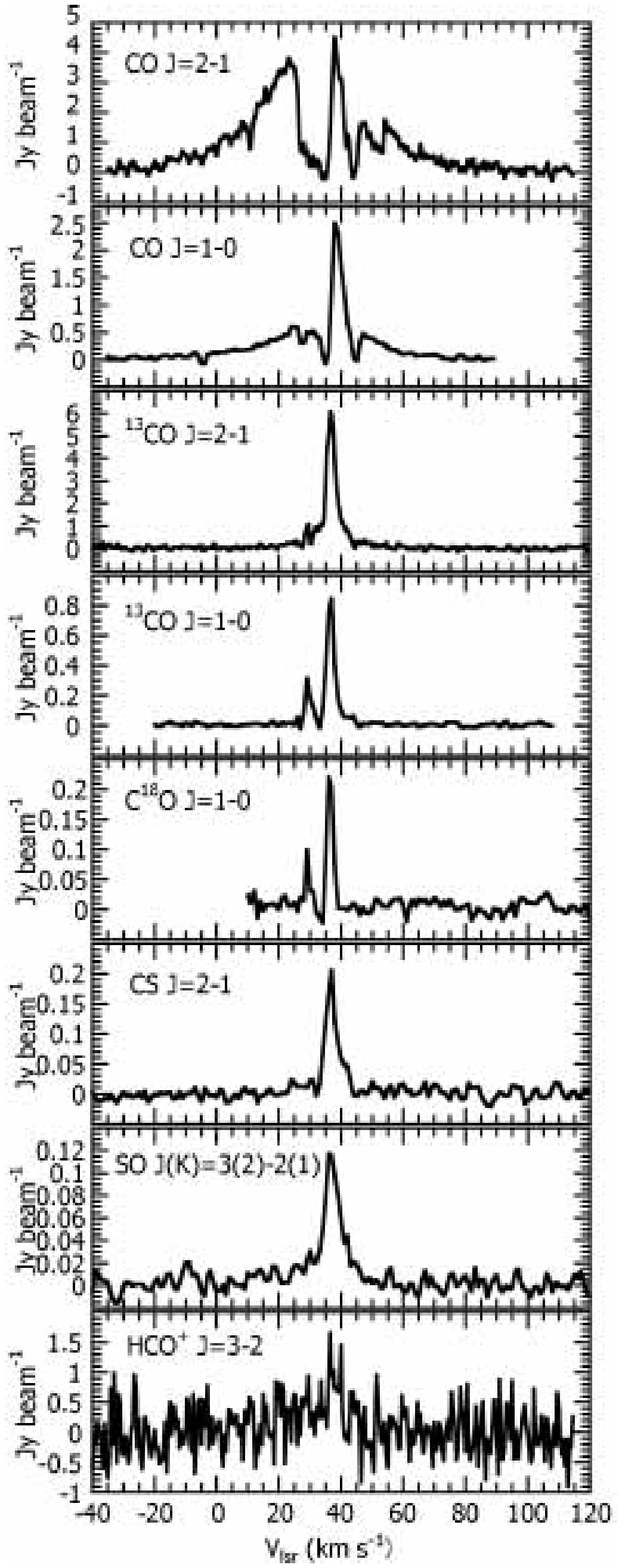}
\caption{Spectra of IRAS 19312+1950 in the CO $J=2$--1, CO $J=1$--0, $^{13}$CO $J=2$--1, $^{13}$CO $J=1$--0, C$^{18}$O $J=1$--0,  CS $J=2$--1, SO $J_{\rm K}=3_{2}$--2$_{1}$ and HCO$^{+}$ $J=3$--2 lines. The spectra were averaged over a circle with a diameter of 15$''$; the averaging circle centered on the mapping center. \label{fig1}}
\end{figure}

\clearpage

\begin{figure}
\epsscale{.80}
\plotone{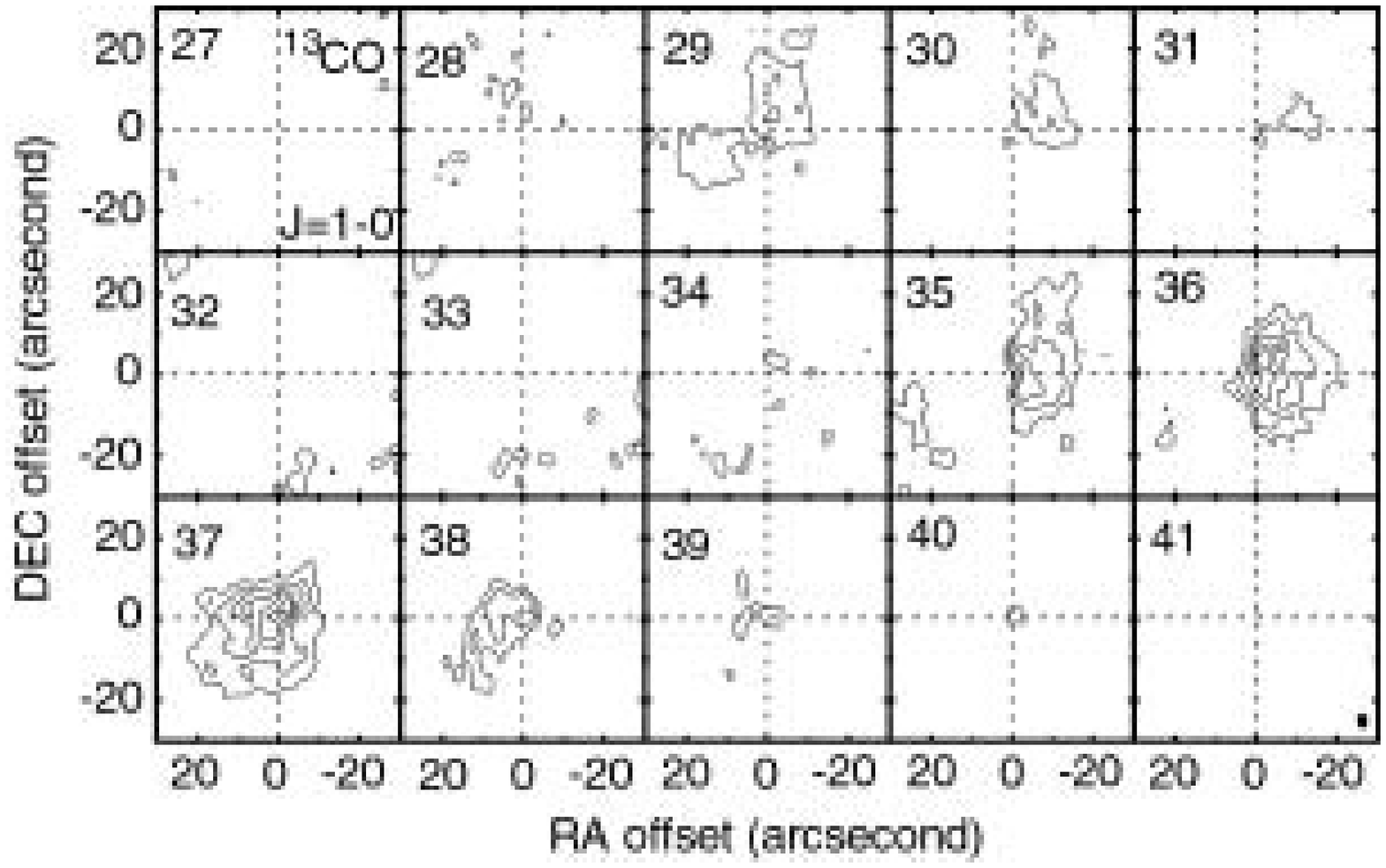}
\caption{Velocity channel maps of the $^{13}$CO $J=1$--0 line. The channel velocities are indicated in the upper-left corners of each panel. The velocity channels are averaged over 1 km s$^{-1}$ intervals. The contours start from a 5 $\sigma$ level, and increase by every 5 $\sigma$. The 1 $\sigma$ level corresponds to $4.96 \times 10^{-2}$ Jy beam$^{-1}$. The synthesized beam is indicated in the lower-right corner. \label{fig2}}
\end{figure}

\clearpage

\begin{figure}
\epsscale{.80}
\plotone{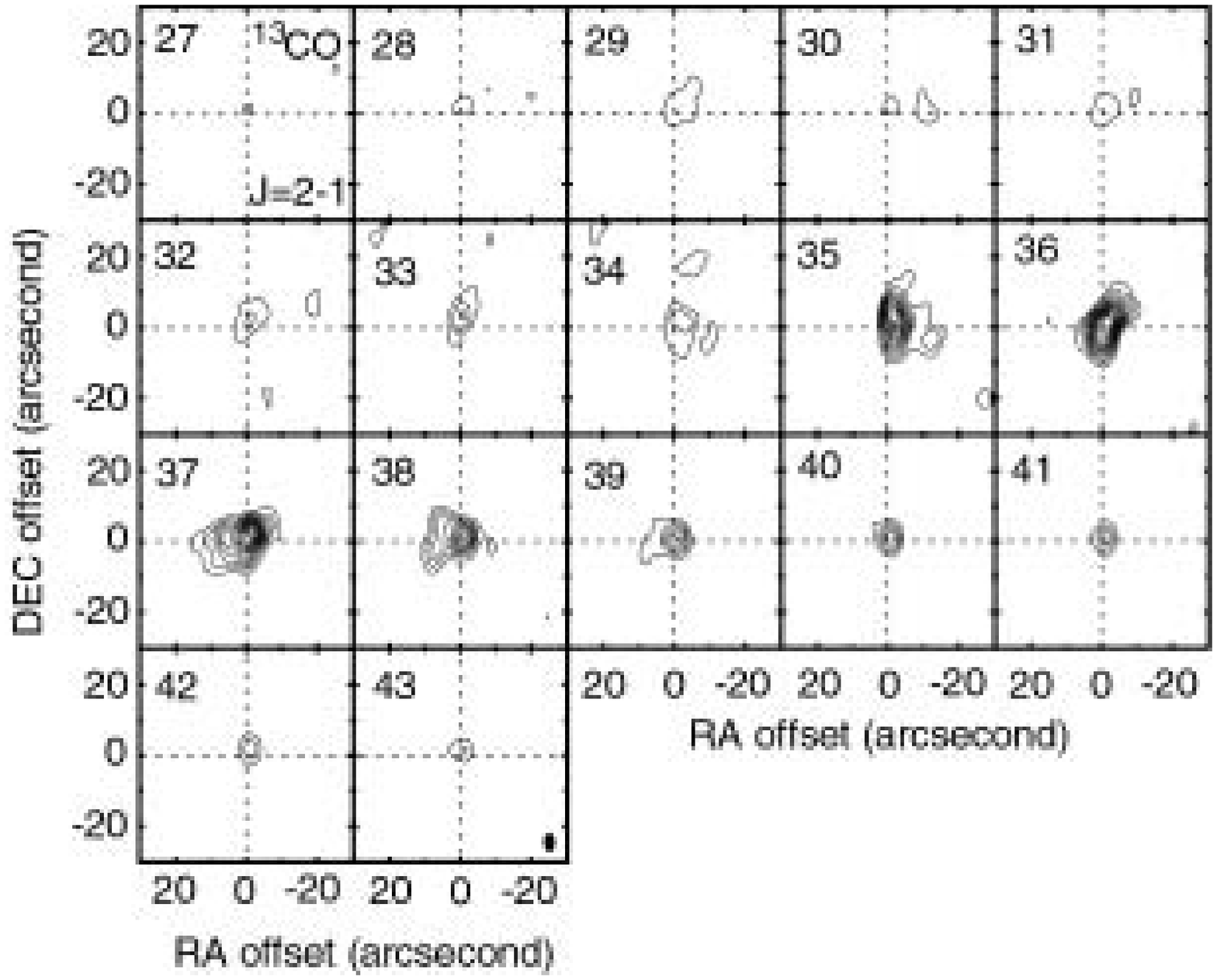}
\caption{Velocity channel maps of the $^{13}$CO $J=2$--1 line. The channel velocities are indicated in the upper-left corners of each panel. The velocity channels are averaged over 1 km s$^{-1}$ intervals. The contours start from a 5 $\sigma$ level, and increase by every 5 $\sigma$. The 1 $\sigma$ level corresponds to $2.24 \times 10^{-1}$ Jy beam$^{-1}$. The synthesized beam is indicated in the lower-right corner. \label{fig3}}
\end{figure}

\clearpage

\begin{figure}
\epsscale{.80}
\plotone{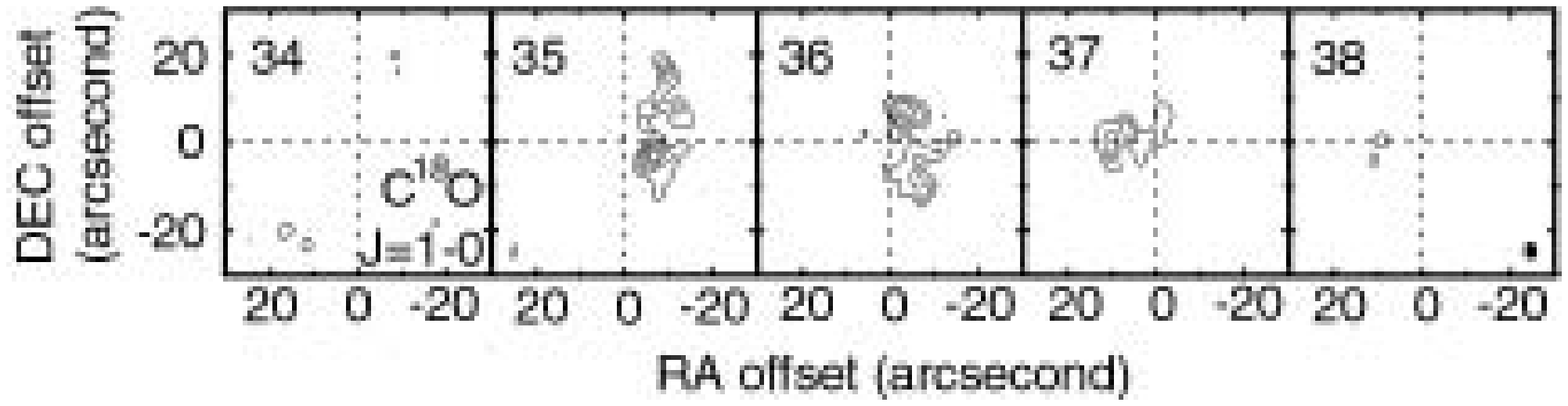}
\caption{Velocity channel maps of the C$^{18}$O $J=1$--0 line. The channel velocities are indicated in the upper-left corners of each panel. The velocity channels are averaged over 1 km s$^{-1}$ intervals. The contours start from a 5 $\sigma$ level, and increase by every 2 $\sigma$. The 1 $\sigma$ level corresponds to $4.66 \times 10^{-2}$ Jy beam$^{-1}$. The synthesized beam is indicated in the lower-right corner. \label{fig4}}
\end{figure}

\clearpage

\begin{figure}
\epsscale{.80}
\plotone{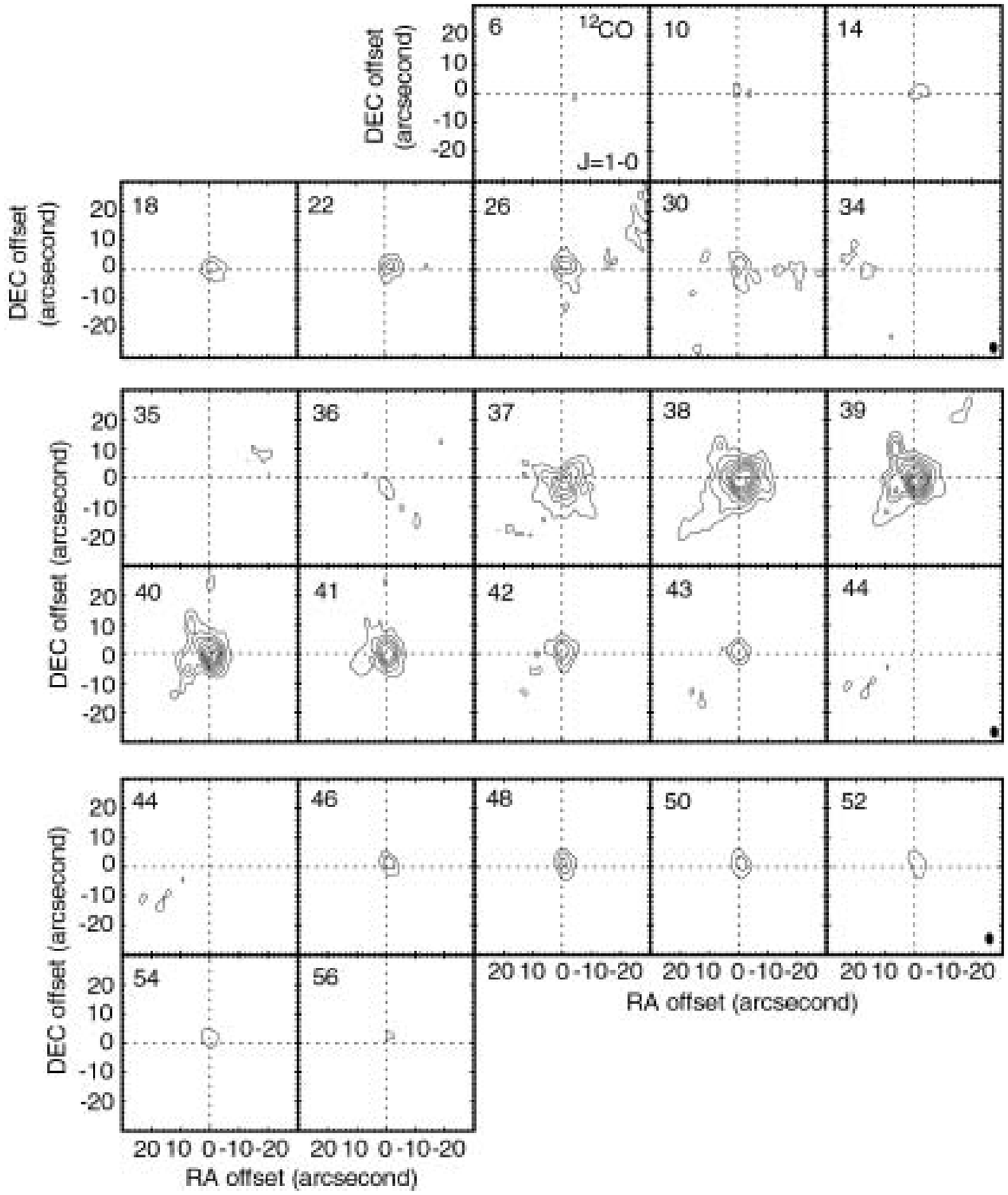}
\caption{Velocity selected channel maps of the $^{12}$CO $J=1$--0 line. The channel velocities are indicated in the upper-left corners of each map. The velocity channels are averaged over 1 km s$^{-1}$ intervals. The contours start from a 5 $\sigma$ level, and increase by every 5 $\sigma$. The 1 $\sigma$ level corresponds to $9.93 \times 10^{-2}$ Jy beam$^{-1}$. The synthesized beam is indicated in the lower-right corners of each panel. The upper, middle and lower panels have different velocity steps: 4, 1 and 2 km s$^{-1}$. \label{fig5}}
\end{figure}

\clearpage

\begin{figure}
\epsscale{.80}
\plotone{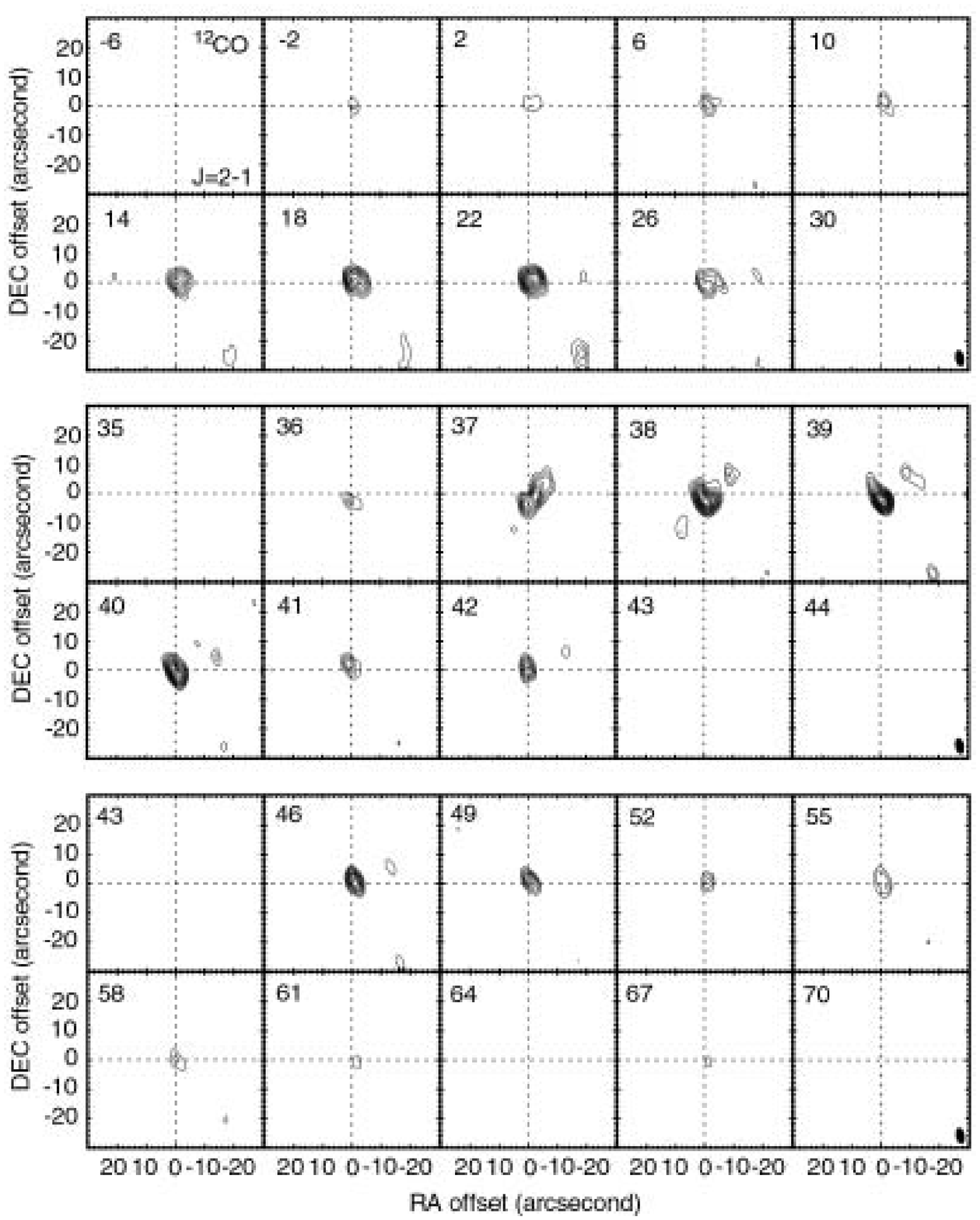}
\caption{Velocity selected channel maps of the $^{12}$CO $J=2$--1 line. The channel velocities are indicated in the upper-left corners of each map. The velocity channels are averaged over 1 km s$^{-1}$ intervals. The contours start from a 5 $\sigma$ level, and increase by every 2 $\sigma$. The 1 $\sigma$ level corresponds to $4.79 \times 10^{-1}$ Jy beam$^{-1}$. The synthesized beam is indicated in the lower-right corners of each panel. The upper, middle and lower panels have different velocity steps: 4, 1 and 3 km s$^{-1}$. \label{fig6}}
\end{figure}

\clearpage

\begin{figure}
\epsscale{.70}
\plotone{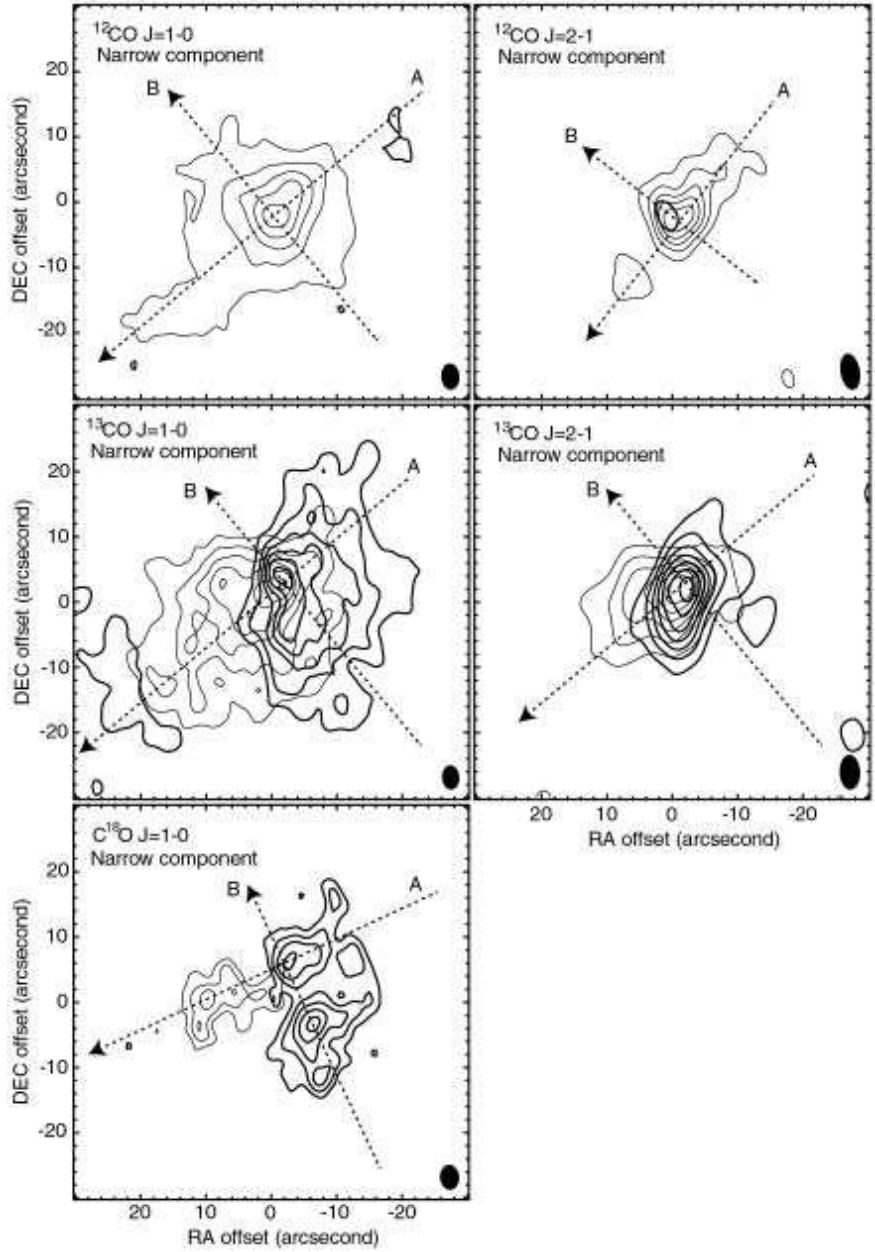}
\caption{Velocity integrated intensity maps of the CO lines (narrow component). The thick and thin contours represent regions in velocity ranges of 35--36 km s$^{-1}$ (thick) and 37--38 km s$^{-1}$ (thin), respectively. The synthesized beams are indicated in the lower-right corners of each panel. The lowest contours correspond to 5$\sigma$ level. The steps of the contours and the 1$\sigma$ rms levels of each data are
10 $\sigma$, $7.02 \times 10^{-2}$ Jy beam$^{-1}$ ($^{12}$CO $J=1$--0), 
5 $\sigma$, $3.39 \times 10^{-1}$ Jy beam$^{-1}$ ($^{12}$CO $J=2$--1), 
5 $\sigma$, $3.51 \times 10^{-2}$ Jy beam$^{-1}$ ($^{13}$CO $J=1$--0), 
10 $\sigma$, $1.59 \times 10^{-1}$ Jy beam$^{-1}$ ($^{13}$CO $J=2$--1), and 
2 $\sigma$, $3.30 \times 10^{-2}$ Jy beam$^{-1}$ (C$^{18}$O $J=1$--0). The dotted arrows ``A'' and ``B'' are cuts used for Figure 12.
\label{fig7}}
\end{figure}

\clearpage

\begin{figure}
\epsscale{.50}
\plotone{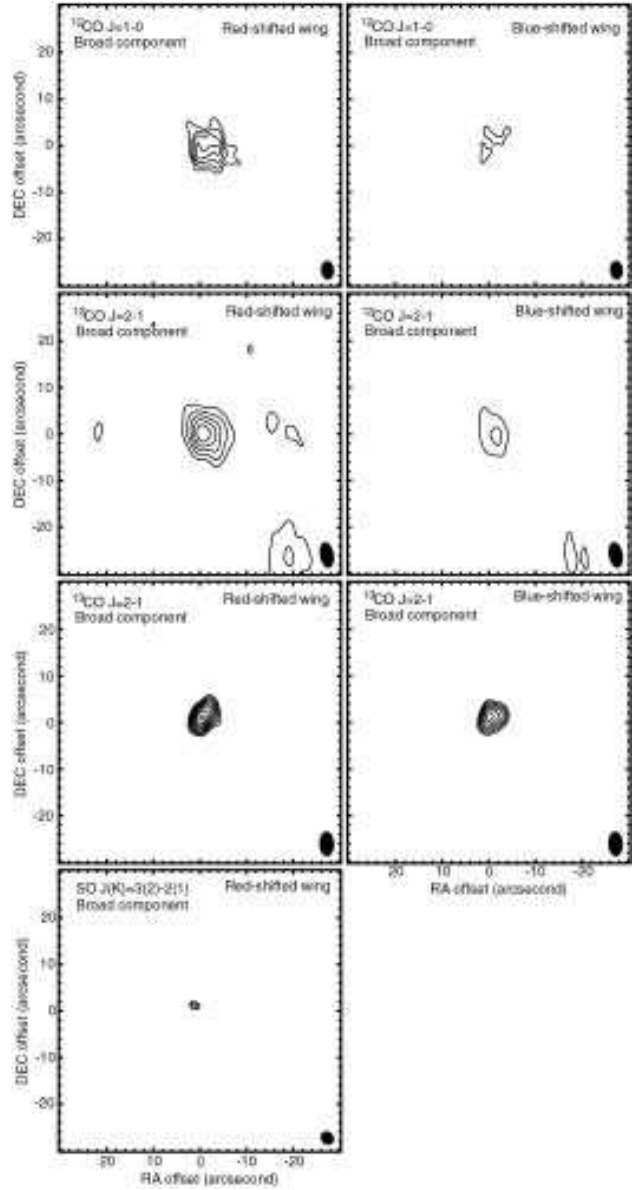}
\caption{Velocity integrated intensity maps of the CO and SO lines (broad component). The left and right columns respectively represent red- and blueshifted wings of the broad component. The integration ranges, corresponding to the blue-  and redshifted wings, are 
$-6$ -- $+10$, $+62$ -- $+79$ km s$^{-1}$ ($^{12}$CO $J=1$--0), 
$-6$ -- $+10$, $+62$ -- $+79$ km s$^{-1}$ ($^{12}$CO $J=2$--1), 
$+21$ -- $+28$, $+45$ -- $+52$ km s$^{-1}$ ($^{13}$CO $J=2$--1), and 
$+21$ -- $+28$, $+45$ -- $+52$ km s$^{-1}$ (SO $J_{\rm K}=3_{2}$--2$_{1}$). 
The synthesized beams are indicated in the lower-right corners of each panel. The lowest contours correspond to 5$\sigma$ level. The steps of contours and 1 $\sigma$ rms levels of each data are 
2 $\sigma$, $2.41 \times 10^{-2}$ Jy beam$^{-1}$ ($^{12}$CO $J=1$--0), 
5 $\sigma$, $1.16 \times 10^{-1}$ Jy beam$^{-1}$ ($^{12}$CO $J=2$--1), 
1 $\sigma$, $8.47 \times 10^{-2}$ Jy beam$^{-1}$ ($^{13}$CO $J=2$--1), and 
0.5 $\sigma$, $1.84 \times 10^{-2}$ Jy beam$^{-1}$ (SO $J_{\rm K}=3_{2}$--2$_{1}$). \label{fig8}}
\end{figure}

\clearpage

\begin{figure}
\epsscale{.80}
\plotone{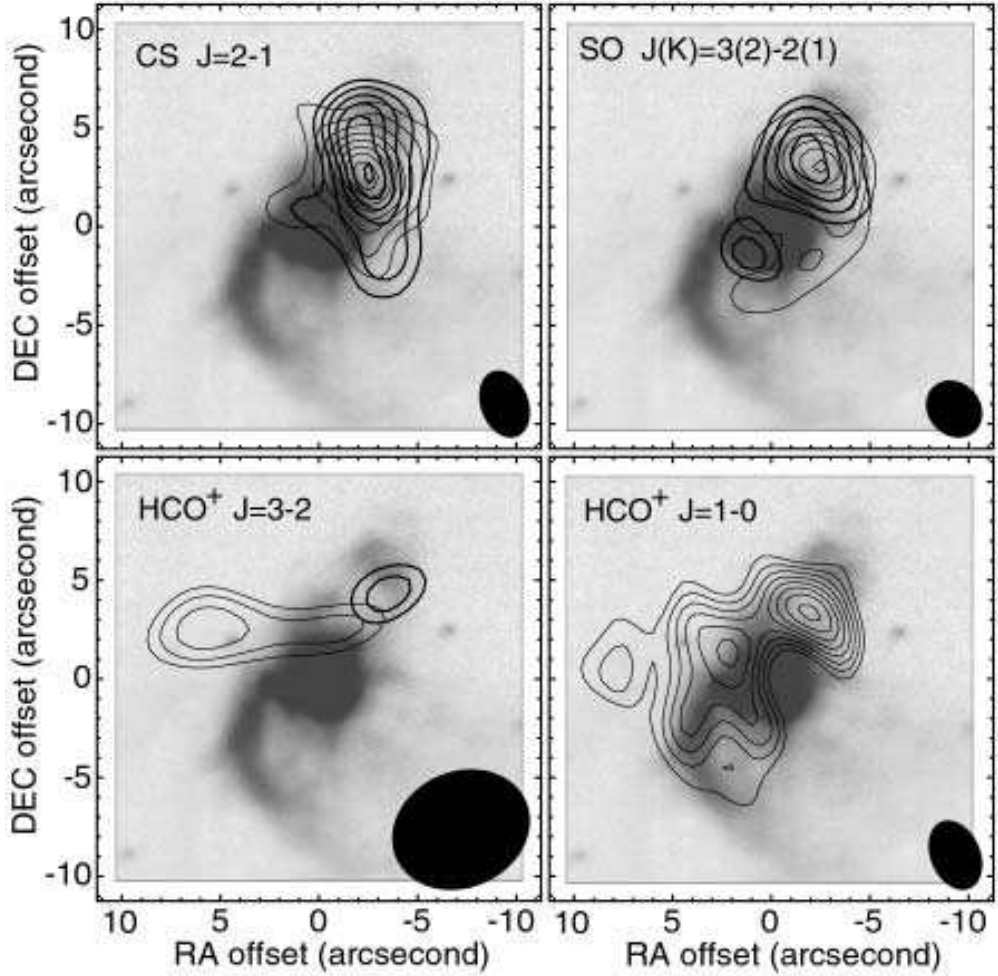}
\caption{Velocity integrated intensity maps of the CS $J=2$--1, SO $J_{\rm K}=3_{2}$--2$_{1}$, HCO$^{+}$ $J=3$--2, and HCO$^{+}$ $J=1$--0 lines (narrow component). The thick and thin contours represent maps in velocity ranges of 34--36 km s$^{-1}$ (thick) and 37--39 km s$^{-1}$ (thin), respectively. The synthesized beams are indicated in the lower-right corners of each panel. The lowest contours correspond to 5$\sigma$ level. Steps of contours and 1 $\sigma$ levels of each line data are 
2 $\sigma$, $3.33 \times 10^{-2}$ Jy beam$^{-1}$ (CS $J=2$--1), 
2 $\sigma$, $3.00 \times 10^{-2}$ Jy beam$^{-1}$ (SO $J_{\rm K}=3_{2}$--2$_{1}$), 
0.5 $\sigma$, $4.80 \times 10^{-1}$ Jy beam$^{-1}$ (HCO$^{+}$ $J=3$--2), and 
1 $\sigma$, $2.20 \times 10^{-2}$ Jy beam$^{-1}$ (HCO$^{+}$ $J=1$--0). 
HCO$^{+}$ $J=1$--0 data are taken from \citet{nak04a}. The background grey scale is a near-infrared composite ($J$, $H$, and $K$ bands) image taken by the CIAO camera on the Subaru telescope (courtesy of K. Murakawa, M. Tamura, and the CIAO group of the National Astronomical Observatory in Japan). \label{fig9}}.
\end{figure}

\clearpage

\begin{figure}
\epsscale{.80}
\plotone{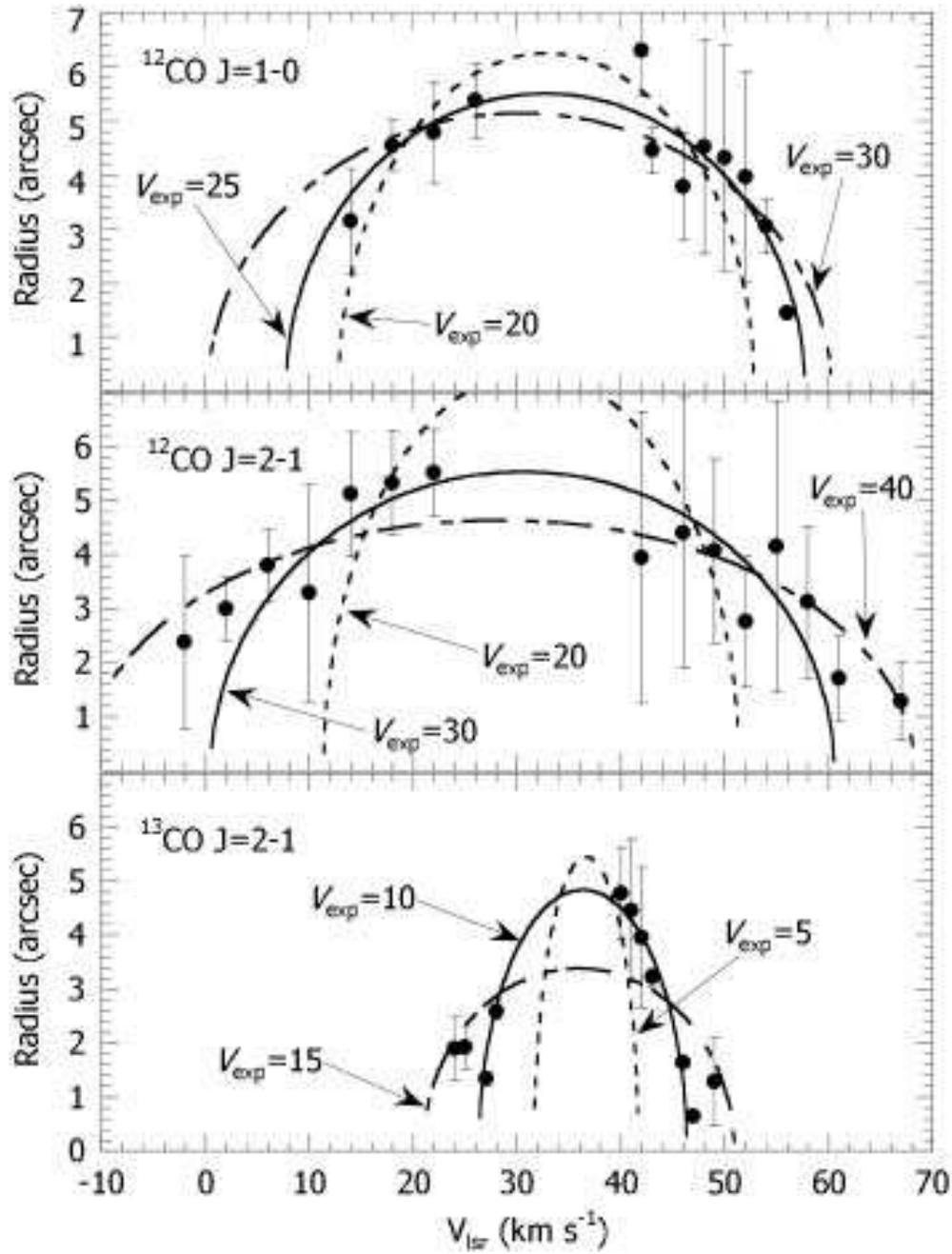}
\caption{Size variation of broad component regions of $^{12}$CO $J=1$--0 (upper), $^{12}$CO $J=2$--1 (middle) and $^{13}$CO $J=2$--1 (lower) as a function of radial velocity. The curves represent models of simple spherical outflows. $V_{\rm exp}$ represents expanding velocities of the models; the unit of the velocity is km s$^{-1}$. \label{fig10}}
\end{figure}

\clearpage

\begin{figure}
\epsscale{.80}
\plotone{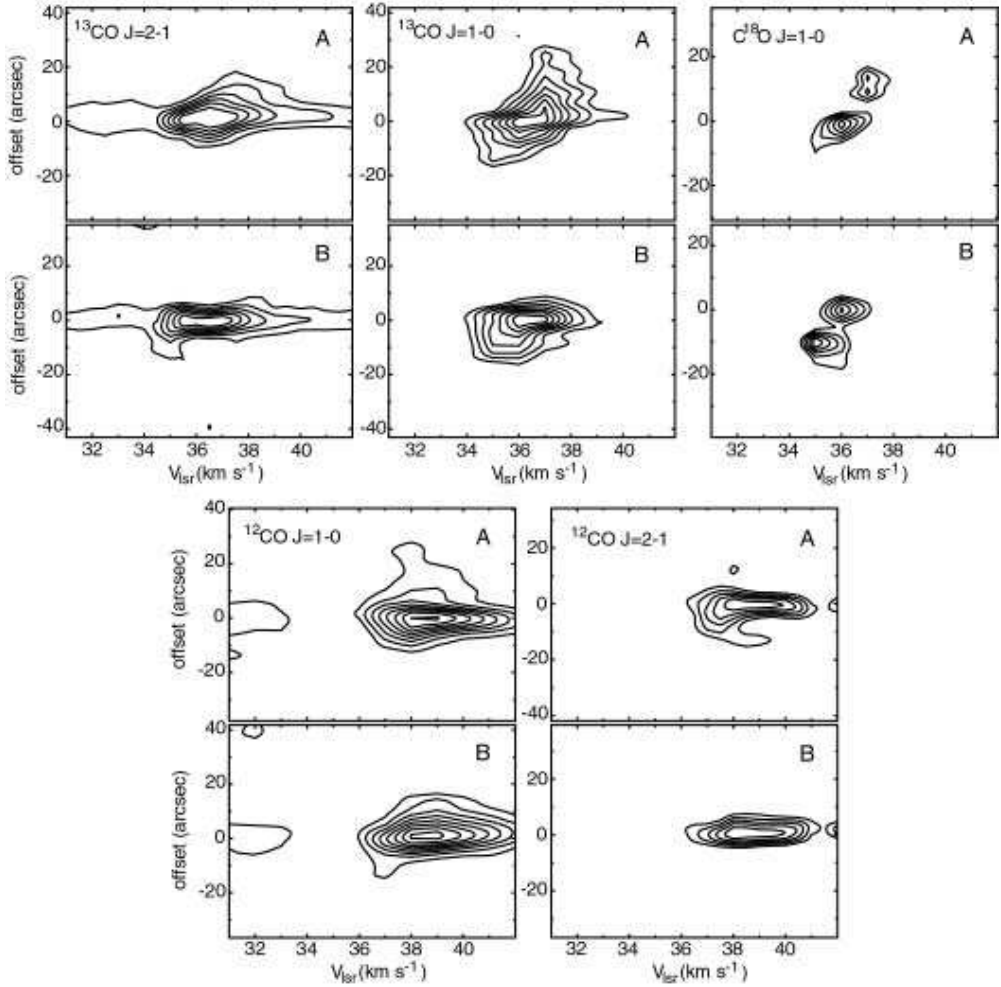}
\caption{Position--velocity maps of CO lines; transitions of lines are indicated in the upper-left corners. ``A'' and ``B'', indicated in the upper-right corners, mean cuts indicated in Figure 7. The cuts A correspond to apparent bipolar axes (see text); the cuts B correspond to perpendicular directions to cuts A. Origins in vertical axes correspond to intersecting points of cuts A and B (see Figure 7). The lowest contours correspond to a 5 $\sigma$ level of rms noises. Contour steps are every 7, 3, 1, 5 and 5 $\sigma$ for $^{13}$CO $J=2$--1, $^{13}$CO $J=1$--0, C$^{18}$O $J=1$--0, $^{12}$CO $J=1$--0 and $^{12}$CO $J=2$--1, respectively. \label{fig11}}
\end{figure}

\clearpage

\begin{deluxetable}{llrrrrrl}
\tablecolumns{8}
\tablewidth{0pc}
\tablecaption{List of Lines Observed}
\tablehead{
\colhead{mol.}    &  \colhead{tran.} &   \colhead{rest freq.}  &
\colhead{rms} & \colhead{int. time} & \colhead{beam} & \colhead{p.a.} &
\colhead{array} \\
\colhead{} & \colhead{}   & \colhead{(GHz)}    & \colhead{(Jy beam$^{-1}$)} &
\colhead{(hr)}    & \colhead{(arcsec)}   & \colhead{(deg)}    & \colhead{}}
\startdata
$^{12}$CO & J=1--0 & 115.271204  & 0.099  & 13.9  & 4.03$\times$2.76 & 3.701 & B, C, D \\
 & J=2--1 & 230.538000  & 0.479  & 12.8  & 5.59$\times$2.96 & 9.947 & C, D \\
$^{13}$CO & J=1--0 & 110.201353  & 0.050  & 12.4  & 3.71$\times$2.64 & 4.038 & B, C, D \\
 & J=2--1 & 220.399000  & 0.224  & 8.7  & 5.33$\times$3.12 & 2.265 & C, D \\
C$^{18}$O & J=1--0 & 109.782160  & 0.047  & 12.4  & 3.84$\times$2.90 & 3.956 & B, C, D \\
CS & J=2--1 & 97.980953  & 0.058  & 5.2  & 3.51$\times$2.41 & 20.29 & B \\
HCO$^{+}$ & J=3--2 & 267.557630  & 0.832  & 11.5  & 7.17$\times$5.88 & $-$66.05 & D \\
SO & J$_{\rm K}$=3$_2$--2$_1$ & 99.299905  & 0.052  & 14.0  & 3.09$\times$2.64 & 41.88 & B \\
\enddata
\end{deluxetable}

\end{document}